\documentclass[aps,prd,twocolumn,superscriptaddress,preprintnumbers,floatfix,nofootinbib,notitlepage,showkeys,showpacs]{revtex4-1}

\usepackage[utf8]{inputenc}

\usepackage{graphicx}
\usepackage{hyperref}
\usepackage{amsmath}
\usepackage{amssymb}
\usepackage{color}

\usepackage{float}
\usepackage[caption=false]{subfig}
\usepackage{bm}

\begin{document}

\title{The Merger Rate of Black Holes in a Primordial Black Hole Cluster}

\author{Viktor D. Stasenko} \email{stasenkovd@gmail.com} \affiliation{Moscow Engineering Physics Institute, National Research Nuclear University MEPhI, 115409 Moscow, Russia}

\author{Alexander A. Kirillov} \email{aakirillov@mephi.ru} \affiliation{Moscow Engineering Physics Institute, National Research Nuclear University MEPhI, 115409 Moscow, Russia}

\begin{abstract}
In this paper, the merger rate of black holes in a cluster of 
primordial black holes (PBHs) is investigated. The clusters have characteristics close to those of typical globular star clusters. A cluster that has a wide mass spectrum ranging from $10^{-2}$ to $10 \, M_{\odot}$ (Solar mass) and contains a massive central black hole of the mass
$M_{\bullet} = 10^3 \, M_{\odot}$ is considered. It is shown that in the process of the evolution of cluster, the merger rate changed significantly, and by now, the PBH clusters have passed the stage of active merging of the black holes inside~them.
\end{abstract}

\maketitle

\section{Introduction}
Since the registration of gravitational waves from black hole mergers, interest in primordial black holes (PBHs) and their clusters has increased. Models of the formation of PBH clusters  are considered with a narrow mass spectrum \cite{2006PhRvD..73h3504C, PhysRevD.100.123544, PhysRevD.98.123533} and a wide one \cite{2000hep.ph....5271R, 2001JETP...92..921R, 2005APh....23..265K}. The first type of clusters appears in the early Universe due to the collapse of density perturbations. The second type describes cluster formation via gravitational collapse of domain walls appearing in the early Universe as a result of the evolution of scalar fields  in the inflationary epoch; see review \cite{2019EPJC...79..246B} and references therein. A cluster is formed after detachment from the Hubble flow and consequent collapse of domain walls into PBHs. Finally, it forms at the redshift $z \sim 10^{4}$ and obtains characteristics close to those of globular star clusters. Despite the similarities, the considered PBHs clusters might have black holes with a wide range of masses and a massive black hole at the center. We discussed the evolution of these systems in Refs. \cite{Stasenko_2020, 2020arXiv201103099S}  in the framework of two-body relaxation; however, the question of the black hole merger rate was not studied. This paper continues the previous investigation and explores the merger rate of black holes during the evolution of such clusters.

There are two possibilities: mergers of black holes with each other and with a massive central black hole (CBH). The first possibility was discussed in Refs. \cite{Ballesteros_2018, 2019EPJC...79..246B, 2020JCAP...11..028D, 2020MNRAS.496..994K, 2021Univ....7...18T} in the context of PBH clusters with narrow mass spectra (with the number of PBH $N_{\text{PBH}} \lesssim 10^4$) and in Refs. \cite{1989ApJ...343..725Q, 1993ApJ...418..147L, 2004ApJ...604..632G, 2020PhRvD.102h3016K, 2019ApJ...873..100C, 2018PhRvD..98l3005R, 2018PhRvL.120o1101R, 2021ApJ...907L..25W, 2020arXiv201203585K} in the context of a dense globular star cluster. The second possibility was considered for globular star clusters \cite{1976ApJ...209..214B, 1977ApJ...211..244L, 1977ApJ...216..883B, 1978ApJ...226.1087C, 1980ApJ...239..685M, 2018PhRvD..98b3021S} and galactic nuclei \cite{1991ApJ...370...60M, 2005PhR...419...65A, 2009ApJ...694..959M,  2015ApJ...804...52M, 2017ApJ...848...10V} in the loss-cone treatment \cite{2013degn.book.....M, 2013CQGra..30x4005M}. In the calculations described in this paper, it is shown that during cluster evolution, mergers mainly occur  with a central black hole; nevertheless, the merger rate is low.

\section{Estimations}

Let us estimate the merger rate of black holes for a single-mass PBH cluster. For the cluster with the total mass $M = 10^5 \, {M}_{\odot}$ (Solar mass) and the characteristic radius $R\sim1$~pc, the characteristic velocity of the PBHs is $v = \sqrt{G M / R} \approx 20$~km~s$^{-1}$ \cite{2019EPJC...79..246B}, where $G$ is the Newtonian gravitational constant. The merger cross-section of two black holes with masses $m$ and $m'$ is given by \cite{1989ApJ...343..725Q, 2002ApJ...566L..17M}:
\begin{equation} 
    \label{cs}
    \sigma = 2 \pi \left ( \frac{85 \pi}{6 \sqrt{2}} \right)^{2/7} \frac{G^2 (m + m')^{10/7} m^{2/7} m'^{2/7} }{c^{10/7} |\textit{\textbf{v}} - \textit{\textbf{v}}'|^{18/7}}.
\end{equation}

This cross-section describes the formation of PBHs of binaries due to gravitational radiation. In the present calculations, it is assumed that the lifetime of a binary is less than the dynamical time of the PBH cluster, the approximation similar to one used in Refs. \cite{1989ApJ...343..725Q, 2020arXiv201203585K, 2020PhRvD.102h3016K}.

The merger rate can be estimated as $\Gamma \approx N v \sigma n$, where $N$ is the number of PBHs in the cluster,  $n$ is the number density of the PBHs. For simplicity, let us consider a cluster with a constant number density of $n = 3 N / 4 \pi R^3$. Thus, the merger rate is: 
\begin{equation}
    \Gamma \sim 10^{-10} \left ( \frac{M}{10^5 \, M_{\odot}} \right)^{17/14} \left ( \frac{R}{1 \, \text{pc}} \right)^{-31/14} \, \text{yr}^{-1}.
\end{equation}

During the cluster evolution, black holes evaporate from the core. This process is a slow diffusion in the velocity space. Therefore, one can assume that black holes escaping the cluster carry away zero energy. Hence, the total energy of the core is conserved. According to the virial theorem, the total energy is $E = - G (m N)^2 /2 R$. Therefore, in the process of BH evaporation, the core size is $R \propto N^2$. As a result, the relationship between the number of black holes in the cluster core and the merger rate 
 is: $\Gamma \propto N^{-3.2}$. 
 
In this study, one considers the clusters with a massive CBH. As a result of the capture of less massive black holes, the CBH mass grows as $\dot{M}_{\bullet} = \rho \sigma_c v$, where $\rho$ is the density of black holes in the cluster, and $\sigma_c = 4 \pi r^2_g (c / v)^2$ is the cross-section for particle capture by a Schwarzschild black hole \cite{1975ctf..book.....L}, $r_g$ is the gravitational radius of the CBH. Estimation of $\dot{M}_{\bullet}$ gives 
\begin{align}
    \dot{M}_{\bullet} & \sim 10^{-11} \left ( \frac{R}{1 \, \text{pc}} \right)^{-5/2} \left ( \frac{M}{10^5 \, M_{\odot}} \right)^{1/2} \nonumber \\
    & \times \left ( \frac{M_{\bullet}}{10^{3} \, M_{\odot}} \right)^2 \, M_{\odot} \, \text{yr}^{-1}.
\end{align}

If the mass of the CBH $M_{\bullet}$ is much less than the core mass, $M_{\rm c}$, then the CBH does not affect the core collapse. Hence, the growth rate of the CBH evolves as $\dot{M}_{\bullet} \propto N^{-4.5}$. Hence, the merger rate of the BHs may play a significant role in both cases. Note that when the core contracts significantly, the CBH stops the collapse and causes an expansion phase~\cite{1977ApJ...217..281S, 1980ApJ...239..685M, 2007PASJ...59L..11H, 2017ApJ...848...10V, 2018PhRvD..98b3021S}.

Note that there is also another mechanism of creating PBH binaries in a cluster based on three-body interaction. The third body can carry away enough energy to form a binary from the two others. The formed binaries lose energy through interactions with single black holes and gravitational radiation. As a result,  pairs of mergers are caused. This possibility was studied for globular star clusters with black holes 
in Refs. \cite{2018PhRvD..98l3005R, 2018PhRvL.120o1101R, 2019ApJ...873..100C, 2020PhRvD.102h3016K, 2020arXiv201203585K, 2021ApJ...907L..25W}. In addition, these binaries might significantly affect the cluster evolution, stop the core collapse, and cause gravothermal oscillations \cite{1975MNRAS.173..729H, 1987ApJ...313..576G, 1989MNRAS.237..757H, 1992PASP..104..981H, 2012MNRAS.425.2493B}. These binaries will descend towards the cluster center under the influence of dynamical friction. In the case considered in this paper, they will be destroyed as a result of interactions with the massive central black hole ($\sim 10^3 \, M_{\odot}$) \cite{1980AJ.....85.1281H, 2020MNRAS.498.4287H}. The characteristic time scale for this process is \cite{2013degn.book.....M}:
\begin{align}
    t_{\text{df}} & \sim \frac{v_{\text{B}}^3}{4 \pi G^2 m_{\text{B}} \rho \ln{\Lambda}} \approx 1.7 \, \left( \frac{v_{\text{B}}}{20 \, \text{km} \, \text{s}^{-1}} \right)^3 \nonumber \\
    & \times \left( \frac{\rho}{10^5 \, M_{\odot} \, \text{pc}^{-1}} \right)^{-1} \left( \frac{m_{\text{B}}}{20 \, M_{\odot}} \right)^{-1} \, \text{Myr},
\end{align} 
where $v_{\text{B}}$ and $m_{\text{B}}$ are the velocity and the mass of the PBH binary, and $\ln{\Lambda} = 10$, where $\ln{\Lambda}$ is the Coulomb logarithm. On the other hand, the merger time $t_\text{gw}$ of two black holes  of the same mass $m$ in a circular orbit with the separation $a$ due to an emission of gravitational waves is given by the expression \cite{1975ctf..book.....L}:
\begin{align}
    t_{\text{gw}} &= \frac{5 c^5 a^4}{512 G^3 m^3} \nonumber \\
    & \approx 1.6 \, \left( \frac{a}{0.01 \, \text{au}} \right)^4 \left( \frac{m}{10 \, M_{\odot}} \right)^{-3} \, \text{Myr}.
\end{align}

The condition $t_{\text{gw}} < t_{\text{df}}$ leads to the maximum value of $a$ for a binary system that merges before its destruction:
\begin{equation}
    a \lesssim 10^{-2} \, \left ( \frac{t_{\text{df}}}{2 \, \text{Myr}} \right)^{1/4} \left ( \frac{m}{10 \, M_{\odot}} \right)^{3/4} \, \text{au}. 
\end{equation}

The probability of the formation  of such hard binaries requires a careful analysis. Thus, we assume that all  binaries that are forming are destroyed by the CBH. The destruction mechanism also remains true if the PBH binaries are primordial in a cluster. These binaries may have a similar origin that of those formed in the early Universe \cite{2017PhRvD..96l3523A, 2017JCAP...09..037R, 2020PhRvD.101d3015V}.

\section{Merger Rate of Black Holes}

The merger rate of black holes with the mass $m$ per unit of phase space, per unit of mass, and per unit of time is given by \cite{1989ApJ...343..725Q}: 
\begin{align} \label{rate1}
    \gamma &= f(\bm{r}, \bm{v}, m) \int \mathrm{d} m' \int \mathrm{d} \bm{v}' \, f(\bm{r}, \bm{v}', m') \nonumber \\
    & \times |\bm{v} - \bm{v}'| \, \sigma\big(m,m',|\bm{v} - \bm{v}'|\big),
\end{align}
where $f$ is the distribution function of black holes in the cluster and $\sigma$ is the merger cross section is defined by \eqref{cs}. The expression \eqref{cs} shows that the cross-section is $\sigma = \sigma(m,m') |\bm{v} - \bm{v}'|^{-18/7}$. In the case of spherical symmetry, Equation \eqref{rate1} reads:
\begin{align}
    \label{rate2}
    \gamma &= \frac{14 \pi f(r,v,m)}{3 v} \int \mathrm{d} m' \, \int \mathrm{d} v' \, v' \sigma(m,m') \nonumber \\
    & \times f(r,v',m') \, \left[ (v + v')^{3/7} - |v - v'|^{3/7} \right].
\end{align}

Let us take the delta-functional approximation for the distribution function of the $i$-th black hole type:
\begin{equation}
    f_{i}(r,v,m) = \frac{n_i(r)}{2 \pi v} \delta \Big( v^2 - \overline{v_{i}^2}(r) \Big) \delta (m - m_{i}),  
\end{equation}
where $\overline{v_{i}^2}(r)$ is the 
mean-squared velocity \cite{2008gady.book.....B}: 
\begin{equation}
    \overline{v_{i}^2}(r) = \frac{4 \pi}{n_{i}(r)} \int\limits_{\phi(r)}^0 \mathrm{d}E \, f_i(E) \Big[2\big(E - \phi(r)\big) \Big]^{3/2},
\end{equation}
and $m_{i}$ and $n_{i}$ are, respectively, the mass and the number density of the $i$-th PBH type and $\phi (r)$ is the gravitational potential of the cluster. In other words, it is assumed that at any point $r$, black holes have the same velocity $\overline{v_{i}^2}(r)$, and their direction is isotropic. Upon integrating in Equation \eqref{rate2}, one obtains the merger rate of the $j$-th BH type with the others:
\begin{align}
\label{mr}
    \Gamma_{j} & = \frac{14 \pi }{3} \sum_i \sigma(m_j, m_i)  \int  \mathrm{d}r \, r^2 \frac{n_{j} n_i}{\overline{v}_{j} \overline{v}_i} \nonumber \\
    & \times \Big[ (\overline{v}_j + \overline{v}_i)^{3/7} - |\overline{v}_j - \overline{v}_i|^{3/7} \Big],
\end{align}
where we introduce the notation $\sqrt{\overline{v^2}} = \overline{v}$. In order to obtain the merger rate of the $j$-th and $i$-th BH types only, one should take only the $i$-th term under the sum in Equation \eqref{mr}.

\section{Results}

To study the evolution of a PBH cluster, the orbit-averaged Fokker--Planck equation with the loss-cone term \cite{2017ApJ...848...10V, Stasenko_2020} is used here. The following form of the density profile is considered:
\begin{equation}\label{rho_cl}
    \rho_i(r) = \rho_{0,i} \left( \frac{r}{r_0} \right)^{-1} \left( 1 + \frac{r^2}{r_0^2} \right)^{-2},
\end{equation}
where $r_0 = 1$~pc, and $\rho_{0,i}$ is the normalization constant. The profile \eqref{rho_cl} is the modified Plummer model and is often used as a toy model for globular star clusters \cite{2008gady.book.....B}. The central cusp $\rho \propto r^{-1}$ is necessary due to the presence of a CBH. It is known that a physically valid density profile in the presence of a CBH must be steeper than $r^{-1/2}$ \cite{2013degn.book.....M}. For convenience, the profile power index is set to $-1$ in the calculations here as soon as the central cusp obeys the Bachkoll--Wolf law, $\rho \propto r^{-7/4}$ \cite {1976ApJ...209..214B} below 1~Myr. Moreover, the results do not depend on the profile shape because the  expansion phase starts very quickly after $\sim 100$~Myr \cite{Stasenko_2020, 2020arXiv201103099S}, the time when the core collapse occurs in the model without a CBH for this cluster.

\begin{figure}
    \includegraphics[width=0.45\textwidth]{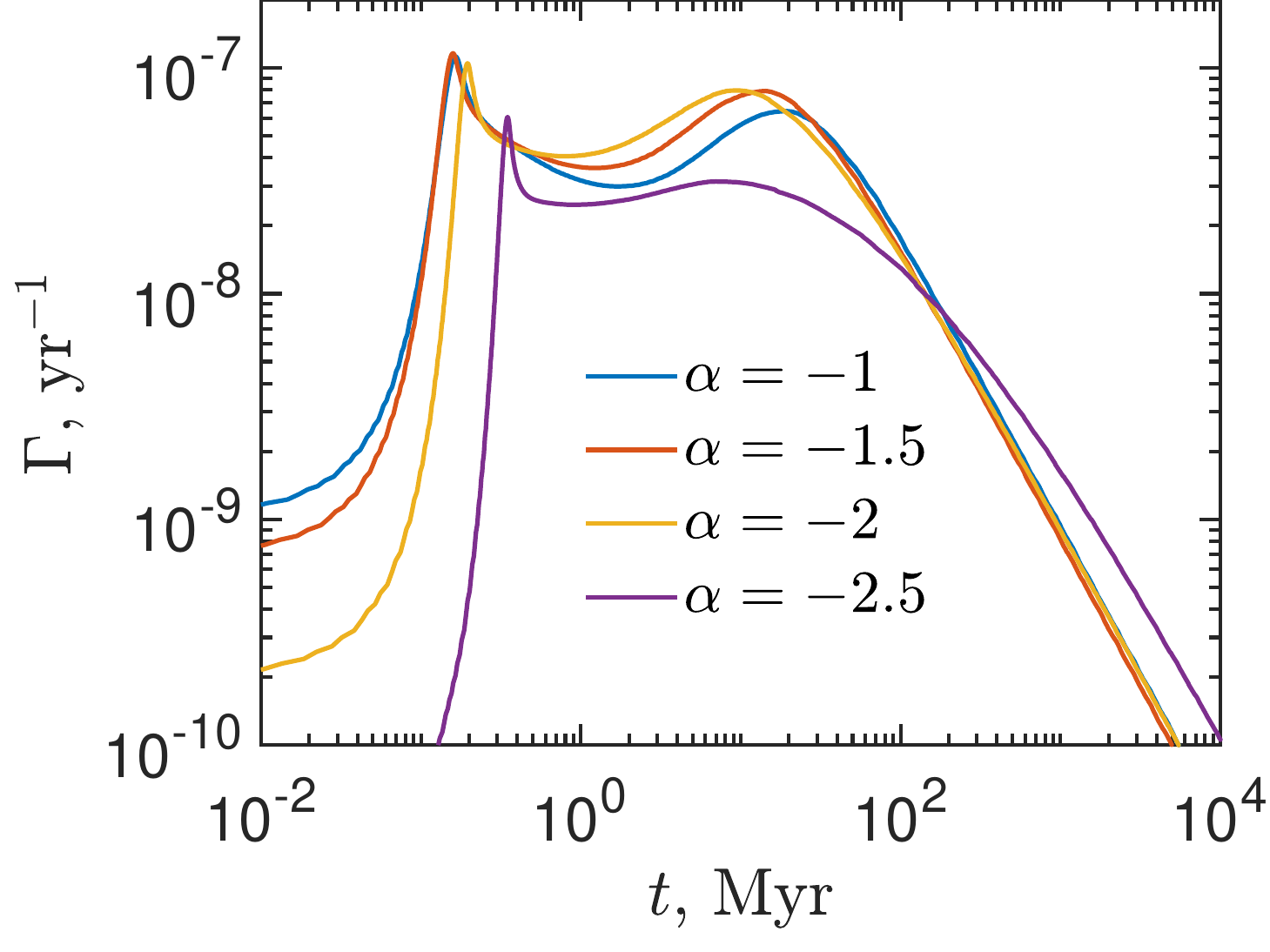}
    \includegraphics[width=0.45\textwidth]{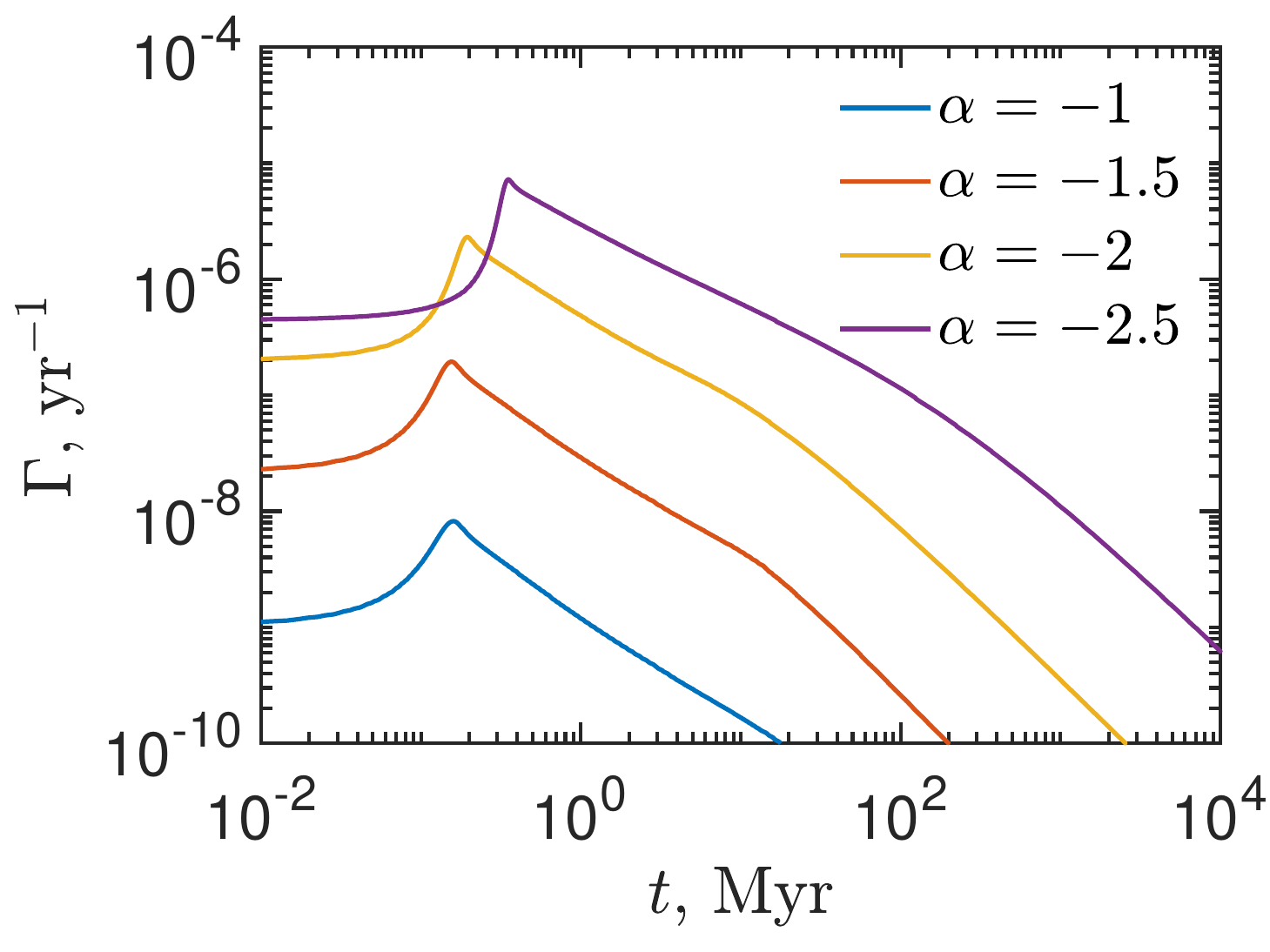}
    \caption{The rate evolution of the capture of black holes  by the central black hole (CBH) for black holes with the mass of $10 \, M_{\odot}$ ({\it top}) and $10^{-2} \, M_{\odot}$ ({\it bottom}). The mass of the CBH is set to $M_{\bullet} = 10^3 \, M_{\odot}$.}
    \label{CBH_rate}
\end{figure}

The mass spectrum of the black holes for the cluster was found to be  \cite{2019EPJC...79..246B}
\begin{equation} 
\label{mass-spectrum}
    \frac{{\rm d} N}{{\rm d} m} \propto m^{\alpha}.
\end{equation}

The black hole masses $m$ range from $10^{-2}$ to $10 \,M_{\odot}$. In the calculations in this paper, the number of mass types of BHs is set to 6, the total mass of the cluster to $10^5 \, M_{\odot}$, and the mass of the CBH to $M_{\bullet} = 10^3 \, M_{\odot}$.

The results of the calculations are shown in Figure~\ref{CBH_rate}, which illustrates the evolution of the merger rate  of different mass types of black holes with the massive central black hole. The initial peak appears due to the Bahcall--Wolf cusp formation \cite{1976ApJ...209..214B}. Further evolution is similar to the core collapse for a cluster without a CBH. After the core collapse time, the cluster slowly expands because the CBH acts as an energy source \cite{1977ApJ...217..281S, 1980ApJ...239..685M, 2017ApJ...848...10V, 2018PhRvD..98b3021S}. In Figure~\ref{CBH_rate} ({\it top}) one can see that, after $\sim$100~Myr, the merger rate of the BHs decreases. For instance, over the next 10~Gyr, the CBH captures $\sim$10 black holes with the mass $m = 10 \, M_{\odot}$ for the model with $\alpha = -1$. The maximum merger number for the period under consideration is $\sim 100$, and it is achieved for the model with $\alpha = -2.5$ and BH masses $m = 10^{-2} \, M_{\odot}$.

\begin{figure}
    \includegraphics[width=0.45\textwidth]{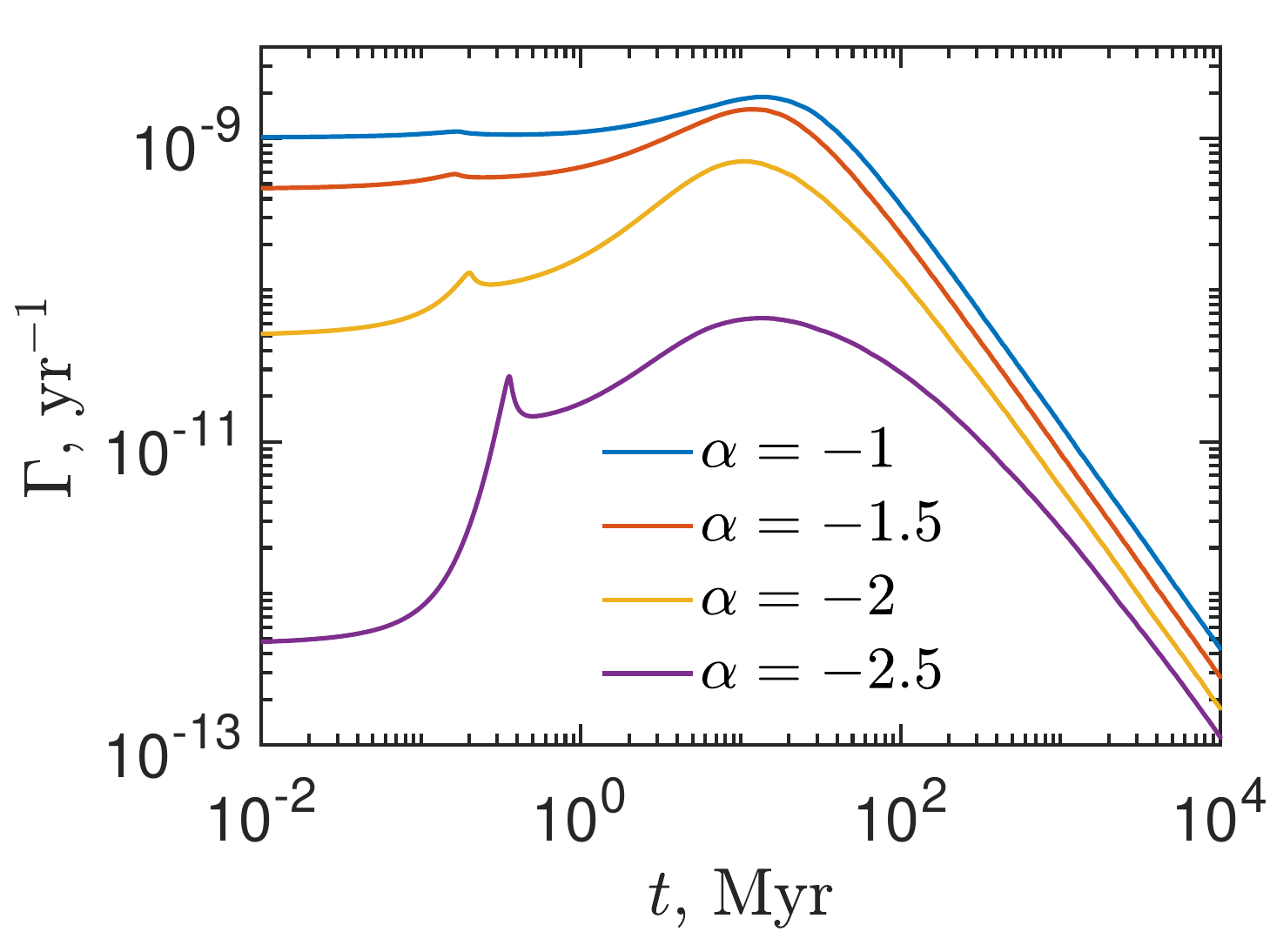}
    \caption{The evolution of the merger rate of the most massive black holes (of the mass of $10\, M_{\odot}$) in the cluster. }
    \label{merge_BH_cbh}
\end{figure}

The evolution of the BH merger rate in the cluster of black hole with no CBH for the different values of $\alpha$ of the mass spectrum \eqref{mass-spectrum} is shown in  Figure~\ref{merge_BH_cbh}. One can see that the merger rate of the black holes with each other is much lower than that with the CBH. Nevertheless, the results can be used to restrict the abundance of PBH clusters with the LIGO/Virgo data \cite{2017PhRvL.118v1101A}. Here, a rough estimate for the clusters in question is given. One can see that the modern merger rate in the cluster is $\Gamma \sim 10^{-13}$~yr$^{-1}$. On the other hand, the merger rate of black holes with the mass $\mathcal{O}(10 \, M_{\odot})$ is estimated by LIGO/Virgo data as $\Gamma \sim 100 $~Gpc$^{-3}$yr$^{-1}$ \cite{2017PhRvL.118v1101A}. Thus, one can obtain the constraints on the number density of the PBH clusters $n_\text{cl} \lesssim 10^{15}$~Gpc$^{-3}$. Taking into account the cluster mass $M_\text{cl} = 10^5$~$M_{\odot}$, one gets $\rho_\text{cl} \lesssim 100 \, M_{\odot}$~kpc$^{-3}$. The critical density has the same order of magnitude: $\rho_c \approx 126 \, M_{\odot}$~kpc$^{-3}$ \cite{2020PTEP.2020h3C01P}. However, an accurate analysis might use more robust constraints.

\section{Discussion}
In this paper, the merger rate evolution of black holes in a primordial black hole (PBH) cluster with a wide mass spectrum and a massive central black hole (CBH) was studied via the Fokker--Planck equation. It was shown that black hole mergers might mainly occur in a typical PBH cluster  due to absorption by a CBH. However, the rapid evolution of the PBH clusters leads to a small merger rate at the present time, compared to the merger rate below 100~Myr. Even the most effective merging process---for the CBH of the mass of $10^{3} \, M_{\odot}$ and black holes of $10 \, M_{\odot}$---only gives the rate $\Gamma \sim 10^{-10}$~yr$^{-1}$. In addition, the merger rate of black holes with masses $\sim 10 \, M_{\odot}$  observed by LIGO/Virgo experiment does not impose restrictions on the abundance of PBH clusters.

This study might serve as a starting point for studying the frequency of gravitational wave signals from mergers of black holes in PBH clusters for the planned LISA experiment \cite{2017arXiv170200786A}. The analysis of the future LISA data can give additional information about the abundance of PBH clusters and/or constrain some models. 

\bibliography{bib}

\end{document}